\documentclass[prb,twocolumn,aps,showpacs]{revtex4}
\usepackage{graphicx}%
\usepackage{dcolumn}
\usepackage{amsmath}
\usepackage{amssymb}
\usepackage{epsfig}

\begin{document}

\title{Bethe-Salpeter equations for the collective-mode spectrum of a superfluid
Fermi gas in a moving optical lattice}

\author{Zlatko G. Koinov}\affiliation{Department of Physics and Astronomy,
University of Texas at San Antonio, San Antonio, TX 78249, USA}
\email{Zlatko.Koinov@utsa.edu} 
\begin{abstract}
We derive the Bethe-Salpeter (BS) equations for the collective-mode
spectrum  of superfluid Fermi gases of equal mixture of atomic Fermi
gas of two hyperfine states loaded into a moving optical lattice. In
a moving lattice the superfluid state is unstable due to spontaneous
emission of the short-wavelength rotonlike excitations which appear
in the spectrum of the collective modes. It is shown that the
spectrum obtained by the BS equations is in an excellent agreement
with the collective-mode dispersion calculated by the perturbation
approach, while there are some differences between the results
obtained by density response function method and by the BS approach.
The difference increases with increasing the lattice velocity, which
can be seen in the analytical approximations for the dispersion
relation in the long-wavelength limit in a weak-coupling regime.
\end{abstract}
\maketitle

\section{Introduction}
The experimental ability to create a superfluid alkali atom Fermi
gas in optical lattices\cite{SFexp, SFexp1} opens a new opportunity
to study strongly correlated quantum many-particle systems and to
emulate high-temperature superconductors. Theoretical description of
ultracold atomic Bose and Fermi gases in optical lattices  has
attracted much
attention.\cite{BEC1,BEC2,BEC3,BEC4,SF1,SF2,SF3,SF4,SF5,SF6,SF7,SF8,SF9,SF10,SF11,SF12,SF13,
ZK,ZK1} Near the Feshbach resonance the atom-atom interaction can be
manipulated in a controllable way because the scattering length
$a_s$ can be changed from the BCS side (negative values) to the BEC
side (positive values) reaching very large values close to
resonance. On BEC  side of the resonance  the spin-up and spin-down
atoms can form diatomic molecules, and  these bosonic molecules can
undergo a BEC at low enough temperature.\cite{BEC1,BEC2,BEC3,BEC4}
In what follow  we focus our attention on the BCS side (negative
scattering length) where BCS superfluidity is expected analogous to
superconductivity. In particular, we consider an equal mixture of
atomic Fermi gas of two hyperfine states with contact interaction
loaded into an optical lattice. The two hyperfine states are
described by pseudospins $\sigma=\uparrow,\downarrow$. We also
assume that the number of atoms in each hyperfine state per site
(the filling factor) is smaller than unity, and that the lattice
potential is sufficiently deep such that the tight-binding
approximation is valid. The system in this case is well described by
the single-band attractive Hubbard model:
\begin{equation}H=-J\sum_{<i,j>,\sigma}\psi^\dag_{i,\sigma}\psi_{j,\sigma}
-\mu\sum_{i,\sigma}\widehat{n}_{i,\sigma}-U\sum_i
\widehat{n}_{i,\uparrow} \widehat{n}_{i,\downarrow}.
\label{Hubb1}\end{equation} Here, the Fermi operator
$\psi^\dag_{i,\sigma}$ ($\psi_{i,\sigma}$) creates (destroys) a
fermion on the lattice site $i$ with pseudospin
$\sigma=\uparrow,\downarrow$ and
$\widehat{n}_{i,\sigma}=\psi^\dag_{i,\sigma}\psi_{i,\sigma}$ is the
density operator on site $i$ with a position vector $\textbf{r}_i$.
$\mu$ is the chemical potential, and the symbol $\sum_{<ij>}$ means
sum over nearest-neighbor sites. $J$ is the tunneling strength of
the atoms between nearest-neighbor sites, and $U$ is the on-site
interaction. On the BCS side the interaction parameter $U$ is
positive (the atomic interaction is attractive). We assume $\hbar=1$
 and the lattice constant $a=1$.

In the case when the periodic array of microtraps is generated by
counter propagating laser beams with different frequencies, the
lattice is not stationary anymore. It is expected that the formation
of a BCS superfluidity is possible, but due to the presence of
supercurrent (or quasimomentum $\textbf{q}=m\textbf{v}$,  where $m$
is the mass of the trapped atoms and  $\textbf{v}$ is the lattice
velocity) the superflow can break down. Recently, the stability of
superfluid Fermi gases loaded into an optical lattice in the
presence of supercurrent has been studied using the second-order
time-dependent perturbation theory\cite{Gam} (this method is a
generalization of the  earlier
 approach by Belkhir and Randeria\cite{BR}), and the Green's function
 formalism.\cite{Yosh1,Yosh} In what follows, we derive the Bethe-Salpeter (BS)
equations for the spectrum of the collective modes in the case of a
moving optical lattice, and compare the BS formalism with the
above-mentioned two methods.

This paper is organized as follows. In Sec. II, we derive the BS
equations for the dispersion of the collective excitations. In Sec.
III, the BS approach is compared with the results obtained by the
perturbation theory, and by the Green's function formalism.
\section{Bethe-Salpeter equations in the presence of supercurrent}
We consider an imbalance mixture of $^6Li$ atomic Fermi gas of two
hyperfine states $|F=1/2, m_f=\pm 1/2>=|F=1/2,
m_f=\uparrow\downarrow>$ with contact interaction loaded into a
moving optical lattice. The total number of $^6Li$  atoms is $M$,
and they are distributed along $N$ sites. The tight-binding form of
the mean-field electron energy is
$\xi(\textbf{k})=2J\left(1-\sum_\nu \cos k_\nu d\right)-\mu$. In the
presence of supercurrent with velocity $\textbf{v}$, the Cooper's
pair has quasimomentum $2m\textbf{v}$.

The Fourier transform of the single-particle Green's function is a
$2\times 2$ matrix $\widehat{G}=\left(
\begin{array}{cc}
G^{\uparrow,\uparrow}&G^{\uparrow,\downarrow}\\
G^{\downarrow,\uparrow} &G^{\downarrow,\downarrow}
\end{array}%
\right)$. In the mean-field approximation the corresponding matrix
elements are as follows:
\begin{equation}\begin{split}
&G^{\uparrow,\uparrow}_\textbf{q}(\textbf{k},\imath\omega_m)=
\frac{u^2_\textbf{q}(\textbf{k})}{\imath\omega_m-\omega_+(\textbf{k},\textbf{q})}+
\frac{v^2_\textbf{q}(\textbf{k})}{\imath\omega_m+\omega_-(\textbf{k},\textbf{q})}\\&
G^{\downarrow,\downarrow}_\textbf{q}(\textbf{k},\imath\omega_m)=
\frac{v^2_\textbf{q}(\textbf{k})}{\imath\omega_m-\omega_+(\textbf{k},\textbf{q})}+
\frac{u^2_\textbf{q}(\textbf{k})}{\imath\omega_m+\omega_-(\textbf{k},\textbf{q})}\\&
G^{\uparrow,\downarrow}_\textbf{q}(\textbf{k},\imath\omega_m)=
G^{\downarrow,\uparrow}_\textbf{q}(\textbf{k},\imath\omega_m)\\&=u_\textbf{q}(\textbf{k})v_\textbf{q}(\textbf{k})\left[
\frac{1}{\imath\omega_m-\omega_+(\textbf{k},\textbf{q})}-
\frac{1}{\imath\omega_m+\omega_-(\textbf{k},\textbf{q})}\right].
\nonumber\end{split}\end{equation} The symbol $\omega_m$ denotes
$\omega_m= (2\pi/\beta)(m +1/2) ;m=0, \pm 1, \pm 2,... $,
$\beta=(k_BT)^{-1}$, $k_B$ is the Boltzmann constant, $T$ is the
temperature. As can be seen, the one-particle excitations in
mean-field approximation are coherent combinations of electronlike
$\omega_+(\textbf{k},\textbf{q})=E(\textbf{k},\textbf{q})+\eta(\textbf{k},\textbf{q})$
and holelike
$\omega_-(\textbf{k},\textbf{q})=E(\textbf{k},\textbf{q})-\eta(\textbf{k},\textbf{q})$
excitations. The coherent factors $u_\textbf{q}(\textbf{k})$ and
$v_\textbf{q}(\textbf{k})$ give the probability amplitudes of these
states in the actual mixture. Here,
$E(\textbf{k},\textbf{q})=\sqrt{\chi^2(\textbf{k},\textbf{q})+\Delta^2}$,
$u_\textbf{q}(\textbf{k})=\sqrt{\frac{1}{2}\left[1+
\frac{\chi(\textbf{k},\textbf{q})}{E(\textbf{k},\textbf{q})}\right]}$,
$v_\textbf{q}(\textbf{k})=\sqrt{\frac{1}{2}\left[1-
\frac{\chi(\textbf{k},\textbf{q})}{E(\textbf{k},\textbf{q})}\right]}$,
and we have used the following notations
$\eta(\textbf{k},\textbf{q})=\frac{1}{2}\left[\xi(\textbf{k}+\textbf{q})-
\xi(\textbf{q}-\textbf{k})\right]$,
$\chi(\textbf{k},\textbf{q})=\frac{1}{2}\left[\xi(\textbf{q}+\textbf{k})+
\xi(\textbf{q}-\textbf{k})\right]$.

 The number and the
gap equations in this case are as follows ($f=M/N$ is the filling
factor):
\begin{equation}
1-f=\frac{1}{N}\sum_{\textbf{k}}\frac{\chi(\textbf{k},\textbf{q})}{E(\textbf{k},\textbf{q})},\quad
1=\frac{U}{N}\sum_{\textbf{k}}\frac{1}{2E(\textbf{k},\textbf{q})}.
\label{NGEqSC}\end{equation}

 The BS equations for the collective mode
$\omega=\omega_\textbf{q}(\textbf{Q})$ and corresponding BS
amplitudes
$$\widehat{\Psi}_\textbf{q}(\textbf{k},\textbf{Q})=
\left(
\begin{array}{c}
\Psi^{\downarrow,\uparrow}_\textbf{q}(\textbf{k},\textbf{Q})\\
\Psi^{\uparrow,\downarrow}_\textbf{q}(\textbf{k},\textbf{Q})
\\
\Psi^{\uparrow,\uparrow}_\textbf{q}(\textbf{k},\textbf{Q})\\
\Psi^{\downarrow,\downarrow}_\textbf{q}(\textbf{k},\textbf{Q})\end{array}%
\right)$$ in the general random phase approximation (GRPA) are as
follows:
\begin{equation}
\widehat{\Psi}_\textbf{q}(\textbf{k},\textbf{Q})=-\frac{U}{2}\widehat{D}\sum_\textbf{p}
\widehat{\Psi}_\textbf{q}(\textbf{p},\textbf{Q})+\frac{U}{2}\widehat{M}\sum_\textbf{p}
\widehat{\Psi}_\textbf{q}(\textbf{p},\textbf{Q}),
\nonumber\end{equation} The first and second terms in the last
equation represent  the direct and exchange
interactions:\begin{widetext}
\begin{equation}
\widehat{D}=\left(
\begin{array}{cccc}
K_\textbf{q}^{\left(\downarrow,\downarrow,\uparrow,\uparrow\right)}(\textbf{k},\textbf{Q},\imath\omega_p),&
K_\textbf{q}^{\left(\downarrow,\uparrow,\downarrow,\uparrow\right)}(\textbf{k},\textbf{Q},\imath\omega_p)&0&0\\
K_\textbf{q}^{\left(\uparrow,\downarrow,\uparrow,\downarrow\right)}(\textbf{k},\textbf{Q},\imath\omega_p),&
K_\textbf{q}^{\left(\uparrow,\uparrow,\downarrow,\downarrow\right)}(\textbf{k},\textbf{Q},\imath\omega_p)&0&0\\
K_\textbf{q}^{\left(\uparrow,\downarrow,\uparrow,\uparrow\right)}(\textbf{k},\textbf{Q},\imath\omega_p),&
K_\textbf{q}^{\left(\uparrow,\uparrow,\downarrow,\uparrow\right)}(\textbf{k},\textbf{Q},\imath\omega_p)&0&0\\
K_\textbf{q}^{\left(\downarrow,\downarrow,\uparrow,\downarrow\right)}(\textbf{k},\textbf{Q},\imath\omega_p),&
K_\textbf{q}^{\left(\downarrow,\uparrow,\downarrow,\downarrow\right)}(\textbf{k},\textbf{Q},\imath\omega_p)&0&0
\end{array}%
\right), \widehat{M}=\left(
\begin{array}{cccc}
0&0&K_\textbf{q}^{\left(\downarrow,\downarrow,\downarrow,\uparrow\right)}(\textbf{k},\textbf{Q},\imath\omega_p),&
K_\textbf{q}^{\left(\downarrow,\uparrow,\uparrow,\uparrow\right)}(\textbf{k},\textbf{Q},\imath\omega_p)\\
0&0&K_\textbf{q}^{\left(\uparrow,\downarrow,\downarrow,\downarrow\right)}(\textbf{k},\textbf{Q},\imath\omega_p),&
K_\textbf{q}^{\left(\uparrow,\uparrow,\uparrow,\downarrow\right)}(\textbf{k},\textbf{Q},\imath\omega_p)\\
0&0&K_\textbf{q}^{\left(\uparrow,\downarrow,\downarrow,\uparrow\right)}(\textbf{k},\textbf{Q},\imath\omega_p),&
K_\textbf{q}^{\left(\uparrow,\uparrow,\uparrow,\uparrow\right)}(\textbf{k},\textbf{Q},\imath\omega_p)\\
0&0&K_\textbf{q}^{\left(\downarrow,\downarrow,\downarrow,\downarrow\right)}(\textbf{k},\textbf{Q},\imath\omega_p),&
K_\textbf{q}^{\left(\downarrow,\uparrow,\uparrow,\downarrow\right)}(\textbf{k},\textbf{Q},\imath\omega_p)
\end{array}%
\right).
 \nonumber\end{equation}
Here, $\omega_{p}=(2\pi/ \beta)p ; p=0, \pm 1, \pm 2,...$ is a Bose
frequency, and we have introduced the two-particle propagator
$K_\textbf{q}^{\left(i,j,k,l\right)}$ ($
i,j,k,l=\{\uparrow,\downarrow\}$):
\begin{equation}
K_\textbf{q}^{\left(i,j,k,l\right)}(\textbf{k},\textbf{Q},\imath\omega_l)=
\sum_{\omega_m}G_\textbf{q}^{i,j}(\textbf{k}+\textbf{Q};\imath\omega_p+\imath\omega_m)
G_\textbf{q}^{k,l}(\textbf{k};\imath\omega_m)
\nonumber\end{equation} As in the case of a stationary
lattice,\cite{ZK,ZK1} the BS equations can be reduced to a set of
two equations. At zero temperature, the corresponding equations are
as follows:
\begin{equation}\begin{split}&\left[\omega+\Omega_\textbf{q}(\textbf{k},\textbf{Q})-
\varepsilon_\textbf{q}(\textbf{k},\textbf{Q})\right]
G_\textbf{q}^{+}(\textbf{k},\textbf{Q})=
\frac{U}{2N}\sum_{\textbf{p}}
\left[\gamma^\textbf{q}_{\textbf{k},\textbf{Q}}\gamma^\textbf{q}_{\textbf{p},\textbf{Q}}+
l^\textbf{q}_{\textbf{k},\textbf{Q}}l^\textbf{q}_{\textbf{p},\textbf{Q}}\right]
G_\textbf{q}^{+}(\textbf{p},\textbf{Q})\\&-\frac{U}{2N}\sum_{\textbf{p}}
\left[\gamma^\textbf{q}_{\textbf{k},\textbf{Q}}\gamma^\textbf{q}_{\textbf{p},\textbf{Q}}-
l^\textbf{q}_{\textbf{k},\textbf{Q}}l^\textbf{q}_{\textbf{p},\textbf{Q}}\right]
G_\textbf{q}^{-}(\textbf{p},\textbf{Q})-\frac{U}{2N}\sum_{\textbf{p}}
\widetilde{\gamma}^\textbf{q}_{\textbf{k},\textbf{Q}}
\widetilde{\gamma}^\textbf{q}_{\textbf{p},\textbf{Q}}
\left(G_\textbf{q}^{+}(\textbf{p},\textbf{Q})
-G_\textbf{q}^{-}(\textbf{p},\textbf{Q})\right)\\&+\frac{U}{2N}\sum_{\textbf{p}}
m^\textbf{q}_{\textbf{k},\textbf{Q}}m^\textbf{q}_{\textbf{p},\textbf{Q}}
\left[G_\textbf{q}^{+}(\textbf{p},\textbf{Q})+G_\textbf{q}^{-}(\textbf{p},\textbf{Q})\right],
\label{NewEq1SC} \end{split}\end{equation}
\begin{equation}\begin{split}
&\left[\omega+\Omega_\textbf{q}(\textbf{k},\textbf{Q})+
\varepsilon_\textbf{q}(\textbf{k},\textbf{Q})\right]
G_\textbf{q}^{-}(\textbf{k},\textbf{Q})=
-\frac{U}{2N}\sum_{\textbf{p}}
\left[\gamma^\textbf{q}_{\textbf{k},\textbf{Q}}\gamma^\textbf{q}_{\textbf{p},\textbf{Q}}+
l^\textbf{q}_{\textbf{k},\textbf{Q}}l^\textbf{q}_{\textbf{p},\textbf{Q}}\right]
G_\textbf{q}^{-}(\textbf{p},\textbf{Q})\\&+\frac{U}{2N}\sum_{\textbf{p}}
\left[\gamma^\textbf{q}_{\textbf{k},\textbf{Q}}\gamma^\textbf{q}_{\textbf{p},\textbf{Q}}-
l^\textbf{q}_{\textbf{k},\textbf{Q}}l^\textbf{q}_{\textbf{p},\textbf{Q}}\right]
G_\textbf{q}^{+}(\textbf{p},\textbf{Q})
-\frac{U}{2N}\sum_{\textbf{p}}
\widetilde{\gamma}^\textbf{q}_{\textbf{k},\textbf{Q}}\widetilde{\gamma}^\textbf{q}_{\textbf{p},\textbf{Q}}
\left(G_\textbf{q}^{+}(\textbf{p},\textbf{Q})
-G_\textbf{q}^{-}(\textbf{p},\textbf{Q})\right)\\&-\frac{U}{2N}\sum_{\textbf{p}}
m^\textbf{q}_{\textbf{k},\textbf{Q}}m^\textbf{q}_{\textbf{p},\textbf{Q}}
\left[G_\textbf{q}^{+}(\textbf{p},\textbf{Q})+G_\textbf{q}^{-}(\textbf{p},\textbf{Q})\right].
\label{NewEq2SC}
\end{split}\end{equation}
 \end{widetext}Here $$\varepsilon_\textbf{q}(\textbf{k},\textbf{Q})=
 E(\textbf{k}+\textbf{Q},\textbf{q})+ E(\textbf{k},\textbf{q}),$$ $$
 \Omega_\textbf{q}(\textbf{k},\textbf{Q})=
 2J\sum_{\nu}\left[\sin(k_{\nu}d)-\sin(Q_{\nu}d+k_{\nu}d)\right]\sin(q_{\nu}d),$$ and the form factors are
defined as follows:
$$\gamma^\textbf{q}_{\textbf{k},\textbf{Q}}=
u^\textbf{q}_{\textbf{k}}u^\textbf{q}_{\textbf{k}+\textbf{Q}}
+v^\textbf{q}_{\textbf{k}}v^\textbf{q}_{\textbf{k}+\textbf{Q}},
l^\textbf{q}_{\textbf{k},\textbf{Q}}=u^\textbf{q}_{\textbf{k}}
u^\textbf{q}_{\textbf{k}+\textbf{Q}}
-v^\textbf{q}_{\textbf{k}}v^\textbf{q}_{\textbf{k}+\textbf{Q}},$$
$$\widetilde{\gamma}^\textbf{q}_{\textbf{k},\textbf{Q}}=
u^\textbf{q}_{\textbf{k}}v^\textbf{q}_{\textbf{k}+\textbf{Q}}
-u^\textbf{q}_{\textbf{k}+\textbf{Q}}v^\textbf{q}_{\textbf{k}},
 m^\textbf{q}_{\textbf{k},\textbf{Q}}=
u^\textbf{q}_{\textbf{k}}v^\textbf{q}_{\textbf{k}+\textbf{Q}}+
u^\textbf{q}_{\textbf{k}+\textbf{Q}}v^\textbf{q}_{\textbf{k}}.$$

The BS equations (\ref{NewEq1SC})- (\ref{NewEq2SC}) lead to a set of
four coupled linear homogeneous equations. The existence of a
non-trivial solution requires that the secular determinant
\begin{equation} \left|
\begin{array}{cccc}
U^{-1}+I^\textbf{q}_{\gamma,\gamma}&J^\textbf{q}_{\gamma,l}
&I^\textbf{q}_{\gamma,\widetilde{\gamma}}&J^\textbf{q}_{\gamma,m}\\
J^\textbf{q}_{\gamma,l}&U^{-1}+I^\textbf{q}_{l,l}
&J^\textbf{q}_{l,\widetilde{\gamma}}&I^\textbf{q}_{l,m}\\
I^\textbf{q}_{\gamma,\widetilde{\gamma}}&J^\textbf{q}_{l,\widetilde{\gamma}}&
-U^{-1}+I^\textbf{q}_{\widetilde{\gamma},\widetilde{\gamma}}&
J^\textbf{q}_{\widetilde{\gamma},m}\\
J^\textbf{q}_{\gamma,m}&I^\textbf{q}_{l,m}&
J^\textbf{q}_{\widetilde{\gamma},m}&U^{-1}+I^\textbf{q}_{m,m}\end{array}%
\right| \label{BSZ}\end{equation} is equal to zero. Here we have
introduced symbols $I^\textbf{q}_{a,b}$ and $J^\textbf{q}_{a,b}$:
\begin{equation}\begin{split}&
I^\textbf{q}_{a,b}=\frac{1}{N}\sum_{\textbf{k}}\frac{a^\textbf{q}_{\mathbf{k}%
,\mathbf{Q}}b^\textbf{q}_{\mathbf{k}%
,\mathbf{Q}}\varepsilon_\textbf{q}(\textbf{k},\textbf{Q})}
{\left[\omega+\Omega_\textbf{q}(\textbf{k},\textbf{Q})\right]^2-\varepsilon^2_\textbf{q}(\textbf{k},\textbf{Q})}
,\\&
J^\textbf{q}_{a,b}=\frac{1}{N}\sum_{\textbf{k}}\frac{a^\textbf{q}_{\mathbf{k}%
,\mathbf{Q}}b^\textbf{q}_{\mathbf{k}%
,\mathbf{Q}}\left[\omega+\Omega_\textbf{q}(\textbf{k},\textbf{Q})\right]}
{\left[\omega+\Omega_\textbf{q}(\textbf{k},\textbf{Q})\right]^2-\varepsilon^2_\textbf{q}(\textbf{k},\textbf{Q})}.
\label{JSC}\end{split}\end{equation}
 The quantities
$a^\textbf{q}_{\mathbf{k}%
,\mathbf{Q}}$ and $b^\textbf{q}_{\mathbf{k}%
,\mathbf{Q}}$ could be one of the
four form factors: $l^\textbf{q}_{\mathbf{k},\mathbf{Q}},m^\textbf{q}_{%
\mathbf{k},\mathbf{Q}},\gamma^\textbf{q}_{\mathbf{k},\mathbf{Q}}$ or $\widetilde{%
\gamma }^\textbf{q}_{\mathbf{k},\mathbf{Q}}$.

The secular determinant (\ref{BSZ}) is the main result in our BS
approach. This determinant provides the spectrum of the collective
excitations (in the particle and spin channels) $\omega(\textbf{Q})$
in a uniform manner.
\begin{center}
\begin{figure}[tbp]
\includegraphics[scale=1.1]{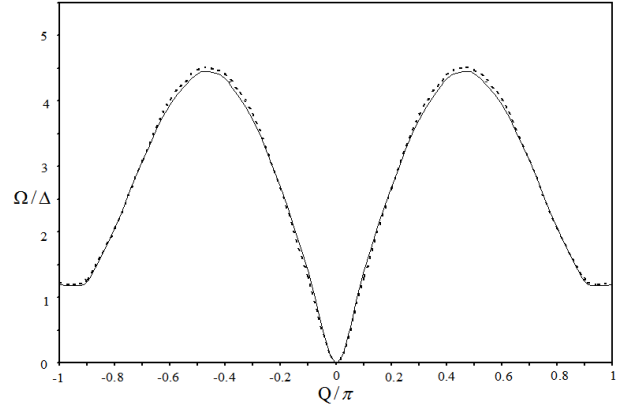} \label{Fig. 1}
\caption{Collective mode energy in a stationary 1D optical lattice
at coupling $U=2J$. The filling factor, the gap and the chemical
potential are $f=0.9$, $\Delta=0.346 J$ and $\mu=1.694 J$,
respectively. The solid line is obtained by the Bethe-Salpeter
method. The dotted line represents the spectrum obtained by the
Green's function method (see FIG. 6a in  Ref.
[\onlinecite{Yosh}])}\end{figure}\end{center}
\section{Comparison with other approaches}
 There exist another two methods that can be used to calculate
 the spectrum of the collective excitations of Hamiltonian
(\ref{Hubb1}). Decades ago, Belkhir and Randeria\cite{BR} calculated
the collective modes by linearizing the Anderson- Rickayzen
equations. In the GRPA the Anderson- Rickayzen equations are reduced
to a set of three coupled equations and the collective-mode spectrum
is obtained by solving the secular equation $Det|\widetilde{A}|=0$,
where
\begin{equation}
\widetilde{A}=\left(
\begin{array}{ccc}
U^{-1}+I_{l,l}&J_{\gamma,l}
&I_{l,m}\\
I_{\gamma,l}&U^{-1}+I_{\gamma,\gamma}
&J_{\gamma,m}\\
I_{l,m}&J_{\gamma,m} &U^{-1}+ I_{m,m}
\end{array}%
\right),\label{Z1}
\end{equation} The quantities $I$ and $J$ are defined by Eq. (\ref{JSC}) when $\textbf{q}=0$.
The perturbation approach by Ganesh et al. provides the following
$3\times 3$ secular determinant (see Eqs. (B8) in Ref.
[\onlinecite{Gam}], where there is a negative sign in front of the
sum in the definition of $\chi_0^{2,3}$):\begin{widetext}
\begin{equation} D=\left(
\begin{array}{ccc}
U^{-1}+I^\textbf{q}_{m,m}&(J^\textbf{q}_{\gamma,m}-I^\textbf{q}_{lm})/2
&-(J^\textbf{q}_{\gamma,m}+I_{l,m})/2\\
I^\textbf{q}_{l,m}-J^\textbf{q}_{\gamma,
m}&U^{-1}+(I^\textbf{q}_{l,l}+
I^\textbf{q}_{\gamma,\gamma}-2J^\textbf{q}_{\gamma,
l})/2&(I^\textbf{q}_{\gamma,\gamma}
-I^\textbf{q}_{l,l})/2\\
-I^\textbf{q}_{l,m}-J^\textbf{q}_{\gamma,m}&(I^\textbf{q}_{\gamma,\gamma}
-I^\textbf{q}_{l,l})/2&U^{-1}+
(I^\textbf{q}_{l,l}+I^\textbf{q}_{\gamma,\gamma}+2J^\textbf{q}_{\gamma,
l})/2
\end{array}%
\right). \label{Gam}\end{equation}\end{widetext}  Since
$Det|D|=det|A|$, where
\begin{equation} Det|A|=\left|
\begin{array}{cccc}
U^{-1}+I^\textbf{q}_{\gamma,\gamma}&J^\textbf{q}_{\gamma,l}
&J^\textbf{q}_{\gamma,m}\\
J^\textbf{q}_{\gamma,l}&U^{-1}+I^\textbf{q}_{l,l}&I^\textbf{q}_{l,m}
\\
J^\textbf{q}_{\gamma,m}
&I^\textbf{q}_{l,m}&U^{-1}+I^\textbf{q}_{m,m}\end{array}\right|,
\label{Josh1}\end{equation} one can say that the perturbation method
is a generalization of the Belkhir and Randeria method to the case
of moving optical lattice.
\begin{center}\begin{figure}[tbp]
\includegraphics{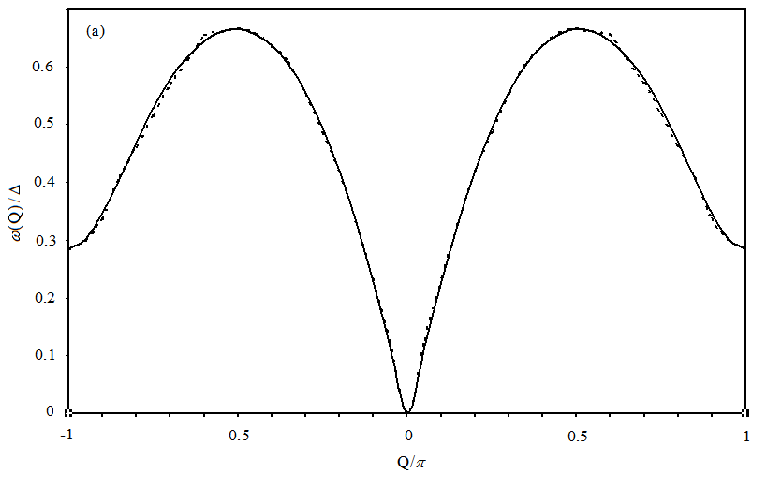} \includegraphics[scale=1.08]{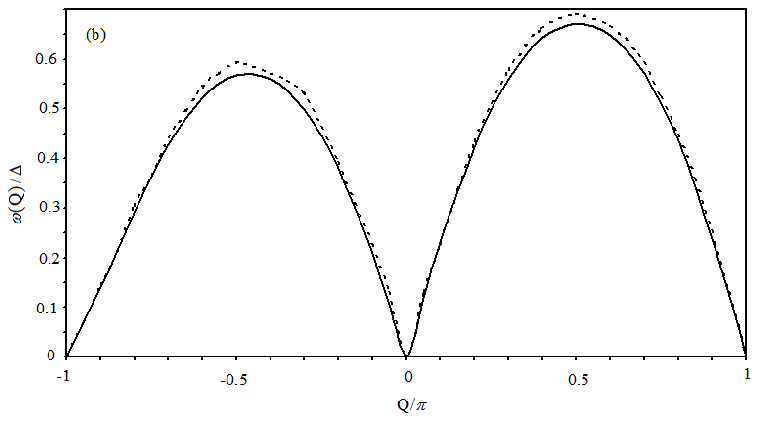} \label{Fig. 2}
\caption{Collective mode energy in a 2D optical lattice at coupling
$U=4.5J$ and filling factor $f=0.8$. The superfluid flows along
$(\pi,\pi)$ direction, and $Q_x=Q_y=Q$. (a) Stationary lattice (the
gap and the chemical potential are $\Delta=1.60 J$ and $\mu=3.32
J$). (b) Moving lattice $q=0.260$ (the gap and the chemical
potential are $\Delta=1.60 J$ and $\mu=3.32 J$). Solid lines are
obtained by the Bethe-Salpeter method while the dotted lines
represent the spectrum obtained by the Green's function method (see
FIG. 11 in  Ref. [\onlinecite{Yosh}]).}\end{figure}
\end{center}
The secular determinant (\ref{BSZ}) can be rearranged as $ \left|
\begin{array}{cccc}A&B^T\\B&C
\end{array}%
\right|=0 $, where  $B^T$ means the transpose matrix, and $B$ and
$C$ are $1\times 3$ and $1\times 1$ blocks, respectively:
\begin{equation} B=\left|
I_{\gamma,\widetilde{\gamma}}\quad J_{l,\widetilde{\gamma}}\quad
J_{\widetilde{\gamma},m} \right|,\quad
C=-U^{-1}+I_{\widetilde{\gamma},\widetilde{\gamma}}.
\label{Z2}\end{equation} Thus, the BS secular equation assumes the
form $Det|A-B^TC^{-1}B|=0$. It should be mentioned that numerical
calculations show that the contributions to $A$ due to the term
$B^TC^{-1}B$ are negligible, and therefore, there is an excellent
agreement between the dispersions obtained by Ganesh et al. in Ref.
[\onlinecite{Gam}] and by the BS secular equation not only in the
case of a stationary lattice (see the conclusion section in Ref.
[\onlinecite{ZK1}]), but in the case of a moving lattice as well.

The method used by Yunomae et al.,\cite{Yosh} known as the Green's
function approach, is a generalization of the C\^{o}t\'{e} and
Griffin\cite{CG1} approach to the collective excitations in s-wave
layered superconductors.
 According to this  approach the spectrum of the collective
excitations of an interacting electron gas (the model includes a
Coulomb  interaction $v(\textbf{r})=e^2/|\textbf{r}|$ and a
short-range attractive interaction $g(\textbf{r})$ between
electrons) is obtained from the poles of the density and spin
response functions in the GRPA. The response functions are obtained
by using the Kadanoff and Baym formalism. The Kadanoff and Baym
formalism is equivalent to the BS formalism. The BS equation for the
two-particle Green's function is $K^{-1}=K^{(0)-1}-I$, where
$K^{(0)}$ is the free two-particle propagator, and the kernel
$I=I_d+I_{exc}$ is a sum of two parts. $I_d$ takes into account the
direct interaction between the electrons, while $I_{exc}$ describes
their exchange interaction. C\^{o}t\'{e} and Griffin have ignored
the long-range Coulomb interaction keeping in $I_d$ only the ladder
diagrams involving the short-range interaction. The exchange
interaction involves bubble diagrams with respect to both unscreened
Coulomb interaction $v(\textbf{r})$ and short-range interaction
attractive  $g(\textbf{r})$. Thus, Eq. (2.33) in Ref.
[\onlinecite{CG1}] is equivalent to the BS equation
$\widetilde{K}=K^{(0)}+K^{(0)}I_d\widetilde{K}$, and Eq. (2.32) in
Ref. [\onlinecite{CG1}] corresponds to the BS equation
$K=\widetilde{K}+\widetilde{K}I_{exc}K$. In other words, the BS
equation $K^{-1}=K^{(0)-1}-I_d-I_{exc}$ is equivalent to both, Eqs.
(2.32) and (2.33) in the paper by C\^{o}t\'{e} and Griffin. The
 kernel of the BS equations  in the case when  $v(\textbf{r})=0$
and $g(\textbf{r})$ replaced by a contact interaction\cite{ZK} does
take into account the direct interaction (terms which depend on
$\gamma$ and $l$) and exchange interaction (terms which depend on
$\widetilde{\gamma}$ and $m$), and therefore, the C\^{o}t\'{e} and
Griffin results in the case of an attractive Hubbard model  should
be similar to the results obtained by the BS approach.  This
statement is supported by FIG. 1 and FIG. 2. The two approaches
provide very similar results, but the difference between them
increases with increasing the lattice velocity.  To illustrate this
statement, we compare the two analytical results for the speed of
sound in a weak-coupling regime in  a moving 1D optical lattice. By
following the calculations that have already been given by Belkhir
and Randeria,\cite{BR} we reduce our secular determinant (\ref{BSZ})
to the following $2\times 2$ determinant:
\begin{equation}
\left|
\begin{array}{cc}
-\frac{Q^2\alpha(\mu^2-4J\mu+\Delta^2)}{4\Delta^2}-\frac{\omega^2\alpha}{4\Delta^2}
&-\frac{\omega\alpha}{2\Delta}\\
-\frac{\omega\alpha}{2\Delta} & 1-\alpha
\end{array}%
\right|=0,\label{DBR1a}
\end{equation} where $\alpha=U/(\pi
\sqrt{4J\mu-\mu^2}$.  The solution $\omega(q)=cQ$ provides the speed
of sound which in a stationary lattice is
$c=\sqrt{1-\alpha}\sqrt{V_F^2-\Delta^2}$, where the Fermi velocity
is $V_F=(a/\hbar)\sqrt{4J\mu-\mu^2}$. The same result has been
obtained in Ref. [\onlinecite{Yosh}]. In the case of a moving
lattice, the corresponding determinant is:
\begin{equation}
\left|
\begin{array}{cc}
A_{11}
&-\frac{\omega\alpha}{2\Delta}+\frac{(2J-\mu)Q\alpha\tan(q)}{2\Delta}\\
-\frac{\omega\alpha}{2\Delta}+\frac{(2J-\mu)Q\alpha\tan(q)}{2\Delta}
& 1-\alpha
\end{array}%
\right|=0.\label{DBR1}
\end{equation}
where $\alpha=U/(\pi \sqrt{4J^2\cos^2(q)-(2J-\mu)^2})$ and
\begin{widetext}
\begin{equation}A_{11}= -\frac{Q^2\alpha(\mu^2-4J\mu+\Delta^2)}{4\Delta^2\cos^2(q)}
-\frac{\omega^2\alpha}{4\Delta^2}
+\frac{(2J-\mu)Q\alpha\omega\tan(q)}{2\Delta^2}-\frac{1}{4}Q^2\tan^2(q)
-\frac{[2J^2-\Delta+4J^2\cos(2q)]Q^2\alpha\tan^2(q)}{2\Delta^2}.
\label{A11}\end{equation} The last determinant provides the
following expression for the phononlike dispersion in the
long-wavelength limit:
\begin{equation}\omega(Q)=(2J-\mu)\tan(q)Qa
+Q\hbar\sqrt{1-\alpha}\sqrt{V^2_F+4J^2\left[2-3\cos^2(q)\right]\tan^2(q)-
\Delta^2\left[1+\frac{1-\alpha}{\alpha}\tan^2(q)\right]}
\label{1D}\end{equation} Here,
$V_F=(a/\hbar)\sqrt{4J^2\cos^2(q)-(2J-\mu)^2}$ is the Fermi velocity
in the presence of a supercurrent. The speed of sound obtained by
the Green's function method is:\cite{Yosh}
\begin{equation}\omega(Q)=(2J-\mu)\tan(q)Qa
+Q\hbar\sqrt{1-\alpha}\sqrt{V^2_F-
\Delta^2\left[1+\frac{1-\alpha}{\alpha^2}\tan^2(q)\right]}
\label{1DY}\end{equation}
\end{widetext}
It should be mentioned that in the case when $\Delta/J<1$, the term
$4J^2\left[2-3\cos^2(q)\right]\tan^2(q)$  is more important than the
term $\Delta^2(1-\alpha)\tan^2(q)/\alpha$ in (\ref{1D}) and the term
$\Delta^2(1-\alpha)\tan^2(q)/\alpha^2$ in (\ref{1DY}).
\section{Conclusion}
We have derived the BS equations for the spectrum of the collective
modes in the case when an equal mixture of atomic Fermi gas of two
hyperfine states with contact interaction is loaded into a moving
optical lattice. It is worth mentioning that there are two
advantages of the BS formalism over the Green's function approach.
First, within the BS approach we obtain the poles of density and
spin response functions in a uniform manner, i.e. one secular
equation provides not only the poles of the density response
function, but the poles of the spin response function as well.
Second, to obtain the poles of the density (spin) response function
within the Green's function method, one has first to take into
account the ladder diagrams with respect to the short-range
interaction, and after that to sum up the bubble diagrams with
respect to both, the short- and long-range interactions.

\end{document}